\documentclass[letterpaper,12pt]{article}   %% LaTeX 2e (preferred)
\usepackage{osajnl2} %% do not use with REVTeX4
\usepackage[draft]{hyperref} %% optional

\usepackage{setspace}
\begin{document}

\title{Spectral modulation for full linear polarimetry
}

\author{Frans Snik,$^{*}$ Theodora Karalidi, and Christoph U. Keller}
\address{Sterrekundig Instituut Utrecht, Princetonplein 5, 3585 CC Utrecht, the Netherlands}
\address{$^*$Corresponding author: f.snik@astro.uu.nl}

\begin{abstract} 
Linear (spectro-)polarimetry is usually performed using separate photon flux measurements after spatial and/or temporal polarization modulation. 
Such classical polarimeters are limited in sensitivity and accuracy by systematic effects and noise. 
We describe a spectral modulation principle which is based on encoding the full linear polarization properties of light in its spectrum. 
Such spectral modulation is obtained with an optical train of an achromatic quarter-wave retarder, an athermal multiple-order retarder and a polarizer. 
The emergent spectral modulation is sinusoidal with its amplitude scaling with the degree of linear polarization and its phase scaling with the angle of linear polarization. 
The large advantage of this passive set-up is that all polarization information is, in principle, contained in a single spectral measurement, thereby eliminating all differential effects that potentially create spurious polarization signals. 
Since the polarization properties are obtained through curve-fitting, the susceptibility to noise is relatively low. 
We provide general design options for a spectral modulator and describe the design of a prototype modulator. 
Currently, the set-up in combination with a dedicated retrieval algorithm is able to measure linear polarization signals with a relative accuracy of 5\%.
\end{abstract}


\maketitle %% null function with osajnl.sty

\section{Introduction}
Polarimetry is a powerful remote-sensing method, but its implementation can be technically challenging.
Classically, the complete information on the linear polarization properties of light is obtained from at least three independent intensity (i.e. photon flux) measurements that are ``filtered'' for a particular polarization state.
As an example, we discuss full linear polarimetry with a four-fold modulation scheme:
\begin{eqnarray}
s_1 &=&t(I+Q)\textrm{;}\nonumber\\
s_2 &=&t(I-Q)\textrm{;}\nonumber\\\
s_3 &=&t(I+U)\textrm{;}\nonumber\\\
s_4 &=&t(I-U)\textrm{,}
\label{modulation}
\end{eqnarray}
with $t$ as the transmission factor, which is ideally identical for all measurements.
The linear Stokes parameters $Q$ and $U$ and the intensity $I$ (all multiplied by the transmission $t$) are trivially obtained from the four intensity measurements $s_{1..4}$.
Usually, the normalized parameters $Q/I$ and $U/I$ are determined, since those are independent of $t$.
The polarization modulation ensures that for each measurement the correct polarization state is transmitted by a polarizer.
Two different classical modulation principles are discerned: spatial and temporal modulation.
In the case of spatial modulation, the light is split into four beams by, e.g., polarizing beamsplitters that select light with linear polarization at $0^{\circ}$ and $90^{\circ}$ (for determining Stokes $Q$ from $s_{1,2}$) and $\pm45^{\circ}$ (Stokes $U$).
With temporal modulation, the measurements $s_{1..4}$ are obtained sequentially with a single beam by, e.g., rotating the polarizer or rotating a half-wave plate in front of a fixed polarizer.
Since the four intensity measurements are independent, polarimetry using spatial or temporal modulation is sensitive to various systematic differential effects as well as to noise.
Spatial modulation requires four different optical paths and four detectors or four different parts of an imaging array to measure the four signals.
Therefore, differential alignment and transmission of the four beams is very critical for a precise polarization measurement.
Also, limited calibration of the gain or flat-field of the imaging array introduces errors in the polarimetry.
On a more practical side, the division into three or four beams significantly increases an instrument's size and mechanical complexity.
As for temporal modulation, differential effects that reduce the polarimeter's performance include time-variations of the source or of the instrument's pointing.
For instance, a remote-sensing instrument that is scanning a source in time as well as modulating polarization in time is prone to spurious polarization signals. 
Also, varying atmospheric properties (such as transmission changes or ``seeing'' due to turbulent refractive index variations) or vibrations of the instrument create such spurious signals.
Temporal modulation requires an active component such as a rotating wave plate or a liquid crystal device, which consume power and may fail at some point during the instrument's life-time.
Many differential errors are eliminated to first order by combining spatial and temporal modulation with the so-called beam-exchange method \cite{Semel}.
Nevertheless, the mechanical complexity of a beam-exchange polarimeter is even larger than for a spatial or temporal modulation polarimeter.
Moreover, all the described methods for polarimetry are sensitive to the random noise in the individual measurements $s_{1..4}$, particularly when the degree of polarization to be measured is very low.
The difference measurements $s_1-s_2$ and $s_3-s_4$ are then usually dominated by random noise.

Here we present a new method for measuring linear polarization as a function of wavelength that is much less sensitive to differential effects.
If the polarization information cannot be stored in the spatial or temporal dimension, and if the instrument is designed to measure the spectral dependency of the light as well (spectropolarimetry), then it is possible to store the polarization information in the spectral dimension.
Such a method, alternately known as channeled spectropolarimetry or polarimetric spectral intensity modulation, was reported independently by Oka \& Kato (1999) \cite{OkaKato}  and Iannarilli et al. (1999)  \cite{Iannarilli,Jones}, but their set-ups measure the complete Stokes vector including circular polarization, whereas our method only applies to linear polarization. 
It is therefore convenient to describe the polarization properties in terms of the degree of linear polarization ($P_L$) and the angle of linear polarization ($\phi_L$), which are obtained from the following coordinate transformation:
\begin{eqnarray}
P_L&=&\frac{\sqrt{Q^2+U^2}}{I}\textrm{;}\\
\phi_L&=&\frac{1}{2}\arctan{\frac{U}{Q}}\textrm{.}
\end{eqnarray}
A ``spectral'' modulation for linear polarization is obtained through encoding the full linear polarization properties into a sinusoidal variation of the intensity spectrum such that the amplitude of the sinusoid scales with $P_L$ and its phase scales with $\phi_L$.
A spectral modulator with such an output can be constructed from standard, passive optical components: a combination of an achromatic quarter-wave retarder, an athermal multiple-order retarder and a polarizer, as is elaborated in sections \ref{principle} and \ref{implementation}. 
The major advantage of this spectral modulation principle is that the measurable polarization properties are, in principle, contained in a single intensity spectrum recording, which eliminates sensitivity to any differential effect.
Moreover, since $P_L$ and $\phi_L$ can be obtained through a curve-fit of the sinusoidal spectral modulation, the susceptibility to noise is lower compared to spatial and temporal modulation.
Of course the implementation of spectral polarization modulation comes at a cost: the spectral resolution of the instrument needs to be much larger than the required spectral resolution of the data-product to carry the additional polarization information in the spectrum.
However, compared to the Oka \& Kato method, which encompasses three sinusoidal modulations with different spectral periodicities, the method described here minimizes the increase in spectral resolution for linear spectropolarimetry.
The Oka \& Kato method is limited by the thermal variation of the employed multiple-order crystals \cite{Okacalibration}.
In section \ref{athermal} we describe a way to create ``athermal'' retarders. 
But even with residual thermal variations of the multiple-order retarder, the measurement of $P_L$ (which is often the key observable) is, in principle, unaffected.

In this paper, we aim to both describe the general principle of spectral polarization modulation (section \ref{principle}), as well as the technical implementation thereof. 
Optical design considerations for the construction of a spectral modulator are discussed in section \ref{implementation}.
We created a dedicated algorithm that successfully disentangles the polarization information from the intensity spectrum.
The algorithm and first results from a prototype spectral modulator are presented in section \ref{retrieval}.
In section \ref{errors} we discuss the systematic instrumental errors pertaining to the retrieved values of $P_L$ and $\phi_L$ and methods to calibrate those.

We foresee a number of applications for full linear spectropolarimetry by spectral modulation.
Any remote-sensing measurement that benefits from the additional value of obtaining the linear polarization properties as well as the spectrum qualifies.
In particular, interesting applications include the characterization of aerosols by analyzing scattered light \cite{aerosols} and the detection of skin cancer \cite{skincancer}, land mines \cite{landmines}, camouflaged objects \cite{camouflage} and of disease in crops and vegetation \cite{crops}.
In all these cases polarization is created or modified by large-angle scattering or reflection of natural or artificial light.
With a spectral modulator, any spectrometer (spectrograph or hyperspectral imager) can be transformed into a spectropolarimeter, albeit at reduced spectral resolution.

%=======

\section{Spectral modulator principle}\label{principle}
\begin{figure}[p]
\centerline{\includegraphics{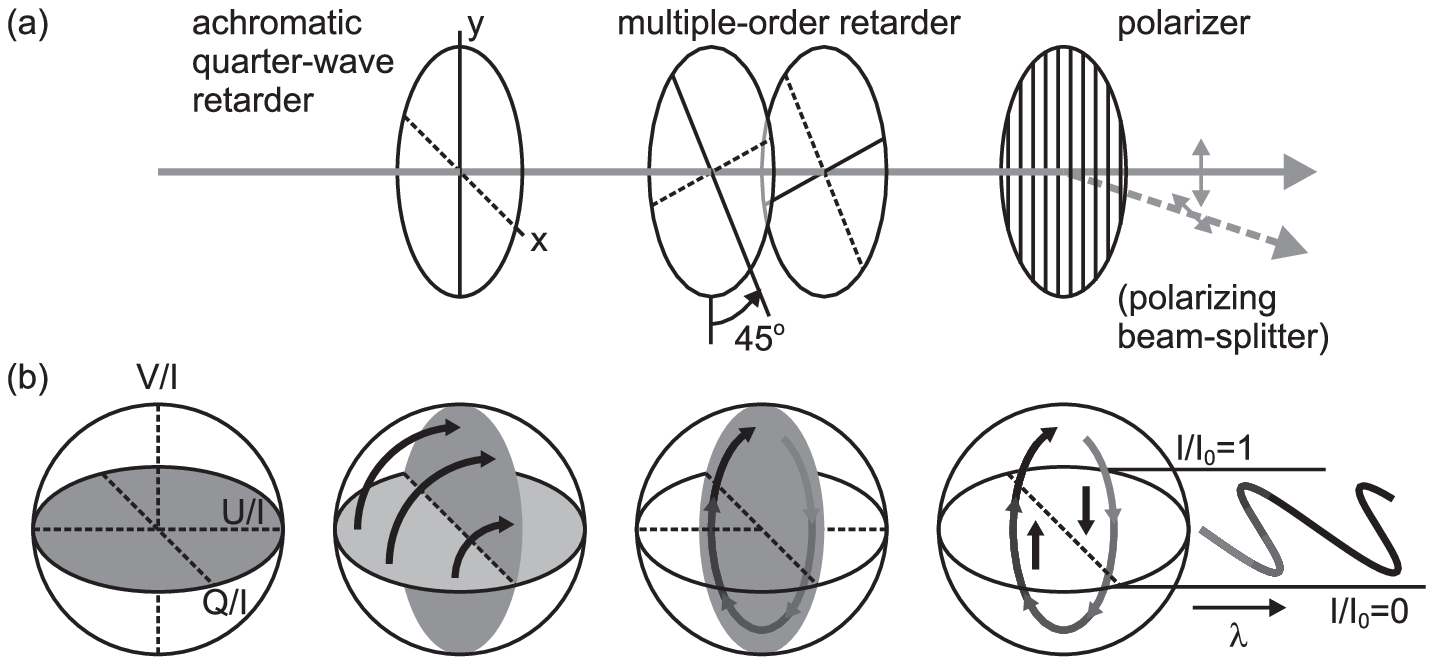}}
\caption{(a) Schematic set-up of the spectral modulator. The solid and dashed lines of the retarders represent the (orthogonal) fast and slow axis respectively. (b) Illustration of the spectral modulation principle on the Poincar\'e sphere. Snik\_AO105143\_fig1.eps}
\label{schematic}
\end{figure}
The spectral modulator for full linear polarimetry creates a sinusoidal variation of the spectrum for which the amplitude scales with the degree of linear polarization ($P_L$) and the phase is determined by the angle of linear polarization ($\phi_L$). 
Such a spectral modulation can be obtained by a combination of an achromatic quarter-wave retarder, an athermal multiple-order retarder and a polarizer.
The schematic set-up including the mutual orientations of the individual components is shown in Fig. \ref{schematic}.a.
The quarter-wave retarder's fast axis and the polarizer are oriented in parallel, although a $90^{\circ}$ rotation of both components does not affect the modulation principle (see below).
The athermal multiple-order retarder has its axes oriented at $\pm45^{\circ}$ from the other two components.
It consists of two crossed wave plates from different crystal materials to make it athermal as explained in section \ref{athermal}.
The retardance of the combined multiple-order crystals is described by $\delta(\lambda,T)$ (see Eq. \ref{delta}).
With the set-up of Fig. \ref{schematic}.a, any spectrum $s_0(\lambda)$ is spectrally modulated as:
\begin{equation}
S(\lambda)=\frac{1}{2}s_0(\lambda)\cdot\Bigg[1\pm P_L(\lambda)\cdot\cos\Bigg(\frac{2\pi\cdot\delta(\lambda,T)}{\lambda}+2\cdot \phi_L(\lambda)\Bigg) \Bigg]\textrm{,}
\label{spectralmodulation}
\end{equation}
with the $\pm$ sign depending on the orientation of the polarizer.
The orientation of $\phi_L$ at the source depends on the $\pm90^{\circ}$ orientation of both retarders.
Note that the modulation is not strictly periodic in wavelength, but in $1/\lambda$ ($\propto$ frequency).
The optimum value of $\delta$ depends on the required spectral resolution.
A typical value for $\delta$ for a spectral modulation period  of $\sim$ 20 nm (which roughly corresponds to the final data-product spectral resolution) in the visible is 25 $\mu$m, as obtained after optimization using the retrieval algorithm introduced in section \ref{retrieval}.
The spectrum $s_0(\lambda)$ is defined as:
\begin{equation}
s_0(\lambda)=I_0(\lambda)t(\lambda)\textrm{,}
\label{transmission}
\end{equation}
where $I_0$ is the intensity spectrum of the light that enters the spectral modulator.
$t(\lambda)$ includes the transmission of the instrument, the detector gain and the analog-to-digital conversion.

The validity of Eq. \ref{spectralmodulation} is easily verified with the Mueller matrix calculus.
As an illustration, Fig. \ref{schematic}.b shows the consecutive operations of the polarization components on the Poincar\'e sphere.
The measurement phase space of linear polarization is spanned by the $(Q/I,U/I)$ plane on which fully polarized light traces out the equator of the sphere and completely unpolarized light is represented by the center of the sphere.
The (achromatic) quarter-wave retarder has its axes in the direction of $\pm Q$ and rotates the $(Q/I,U/I)$ plane to the $(Q/I,V/I)$ plane (with Stokes $V$ representing circular polarization).
The multiple-order retarder has its axes in the $\pm U$ direction and rotates all points in the $(Q/I,V/I)$ plane around the $U/I$ axis.
Since multiple-order crystal retarders are by nature very chromatic, the amount of rotation depends much on wavelength.
Finally, the polarizer in the $\pm Q$ direction makes a projection of all points in the sphere onto the $Q/I$ axis, the position on which corresponds to the relative amount of light (from 0 to 1) that is transmitted by the polarizer.
The action of the multiple-order retarder is thus transformed into the sinusoidal intensity spectrum.
This is essentially the same principle as employed in Lyot filters \cite{Lyot}.
From this illustration it is apparent that the degree of linear polarization $P_L$ directly scales the amplitude of the sinusoidal modulation: fully polarized light creates spectra with maximum modulation amplitude $0\leq I/I_0\leq 1$, whereas completely unpolarized yields a flat spectrum of $I/I_0=\frac{1}{2}$.
A variation of the angle of linear polarization $\phi_L$ is represented by a rotation around the $V/I$ axis of the Poincar\'e sphere.
This rotation is maintained by the action of the achromatic quarter-wave retarder and translated into a phase shift of the sinusoidal modulation by the other two components.
Note that a similar spectral modulation can, in principle, be achieved by replacing the two (linear) retarders by an optically active material (circular retarder).
But in order to create a narrow spectral periodicity of $\sim$ 20 nm it would require unreasonably thick optically active crystals.

The polarizer of the spectral modulator can also be a polarizing beam-splitter.
Eq. \ref{spectralmodulation} implies that the sum of the spectra emerging from a polarizing beam-splitter is ideally equal to the intensity spectrum at full resolution (!), provided that the transmission ratio $t_1(\lambda)/t_2(\lambda)$ (see Eq. \ref{transmission}) between the two beams is known.
Dividing both measured spectra by the intensity spectrum then yields two independent modulation signals scaled as $0\leq s/s_0\leq 1$.
This represents a straightforward method to disentangle the polarization information from the intensity spectrum.
Particularly in cases where narrow features of the intensity spectrum mix up with the polarization modulation, such a dual beam set-up is recommended.
However, if the intensity spectrum is sufficiently smooth, an algorithm can be devised that accurately retrieves the intensity spectrum as well as $P_L$ and $\phi_L$ as described in section \ref{retrieval}, and a single beam set-up suffices.

As with any polarimeter, the modulator is best located as early in the beam as possible in order to minimize the number of optical components that modify the polarization of the source under investigation and/or introduce instrumental polarization.
Ideally, the spectral modulator is located in the entrance pupil of a spectropolarimetric instrument.
The objective lens(es) are then positioned after the polarizer or polarizing beam-splitter to image the source onto the entrance (slit) of the spectrometer.

%=======

\section{Technical implementation}\label{implementation}
The technical implementation for the spectral polarization modulator as schematically depicted in Fig. \ref{schematic}.a depends much on the scientific requirements of a particular instrument.
Different options exist for the optical components based on the required wavelength range, the field-of-view (FOV) and the polarimetric accuracy.
Also, mechanical issues can impose constraints on the optical materials and components (e.g. robustness, clear aperture, complexity).
In this section we discuss different design options for the optical components of a spectral polarization modulator.
We also present our chosen design for a spectral modulator that operates in the extended visible range in combination with a 1 nm resolution slit spectrograph.

\subsection{Achromatic quarter-wave retarder}
Quarter-wave plates based on birefringent crystals are inherently chromatic.
So-called \mbox{(super-)}\-achromatic wave plates can be constructed by using pairs of different crystals \cite{Beckers} or a stack of wave plates at different angles \cite{Pancharatnam} or a combination of both these methods.
Nevertheless, residual wavelength variations of the retardance and of the fast axis angle persist for these wave plates.

\begin{figure}[p]
\centerline{\includegraphics{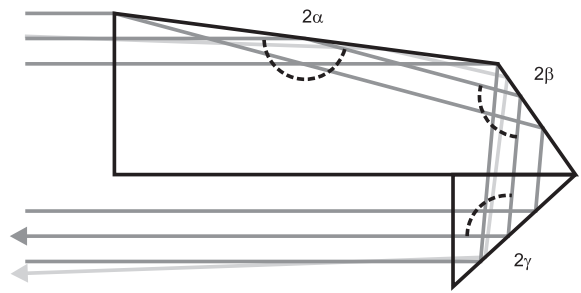}}
\caption{Design of a retroreflecting quarter-wave retarder prism based on 3 TIRs in a fused silica rhomb. The path of light with an angle of 2$^\circ$ from normal incidence is indicated in light grey. Snik\_AO105143\_fig2.eps}
\label{3TIR}
\end{figure}

A second variety of retarders is based on total internal reflections (TIRs) inside glass rhombs for which the retardance due to a single TIR is given by \cite{Keller2002}:
\begin{equation}
\delta_{\textrm{\tiny{TIR}}}=2\arctan\Bigg(-\frac{\cos\theta\sqrt{n(\lambda)\sin^2\theta-1}}{n(\lambda)\sin^2\theta}\Bigg) \textrm{.}
\end{equation}
The major advantage of TIR retarders is the achromaticity of the retardance, which only varies with the wavelength variation of the refractive index $n(\lambda)$.
The most famous example of a quarter-wave retarder based on TIRs is the Fresnel rhomb, which exhibits $\pm2^{\circ}$ retardance variation over the visible range when made from BK7 glass.
Unfortunately, Fresnel rhombs for the visible range cannot be manufactured out of fused silica, which is a favorable material because of its low amount of birefringence induced by internal stresses or by temperature variations \cite{Keller2002}. 
It is possible to tweak the overall retardance of a Fresnel rhomb by means of a coating on the TIR surface \cite{King1966}, but temperature variations can still lead to unacceptable variations of the rhomb's retardance.
A K-prism based on 3 TIRs can be made out of fused silica and has the additional advantage of the lack of lateral beam shift, but it yields a much thicker optical component than a Fresnel rhomb.
The FOV behavior of both a Fresnel rhomb and a K-prism is very anamorphic: the retardance changes by as much as $9\times10^{-2}$ radians for a 1$^{\circ}$ change in beam angle in the plane of the TIRs whereas it is almost constant for beam angle changes in the perpendicular direction.
This behavior does however not severely limit the performance of a spectral modulator in combination with a long-slit spectrograph in the case that the slit is aligned with the ``good'' direction of the Fresnel rhomb or K-prism.
Another type of quarter-wave TIR prism from fused silica can be designed if the output beam direction is allowed to deviate from the input direction as sketched in Fig. \ref{3TIR}.
This way, also the FOV behavior of the prism can be optimized.
For a retroreflecting prism, the equations for determining the three TIR angles $(\alpha,\beta,\gamma)$ are:
\begin{equation}
\theta_c<\alpha,\beta,\gamma<90^\circ \textrm{,}
\label{TIR3-1}
\end{equation}
with $\theta_c$ the critical angle for TIR;
\begin{equation}
\alpha+\beta+\gamma=180^\circ\textrm{;}
\end{equation}
\begin{equation}
\delta_{\textrm{\tiny{TIR}}}(\alpha,\lambda_0)+\delta_{\textrm{\tiny{TIR}}}(\beta,\lambda_0)+\delta_{\textrm{\tiny{TIR}}}(\gamma,\lambda_0)=\frac{\pi}{2}\textrm{;}
\end{equation}
\begin{equation}
\frac{\textrm{d}\delta_{\textrm{\tiny{TIR}}}(\alpha,\lambda_0)}{\textrm{d}\theta}-\frac{\textrm{d}\delta_{\textrm{\tiny{TIR}}}(\beta,\lambda_0)}{\textrm{d}\theta}+\frac{\textrm{d}\delta_{\textrm{\tiny{TIR}}}(\gamma,\lambda_0)}{\textrm{d}\theta}=0 \textrm{.}
\label{TIR3-4}
\end{equation}
The minus sign in the last equation is due to the mirroring effect from the first TIR ($\alpha$; see Fig. \ref{3TIR}).
A solution to Equations \ref{TIR3-1}-\ref{TIR3-4} is found for the case of fused silica with $\alpha=82.5^\circ$, $\beta=50.0^\circ$, $\gamma=47.5^\circ$ at $\lambda_0=$ 500 nm. 
Its optimized FOV yields a retardance variation of $-3\times10^{-2}$ radian for a $\pm2^\circ$ variation of the beam angle in the plane of the rhomb.
Note that variation of the incidence angle causes significant beam shift as indicated in Fig. \ref{3TIR}.

\subsection{Athermal multiple-order retarder}\label{athermal}
The retardance of a crystal wave plate is usually very dependent on temperature, particularly in case of multiple-order retardance.
Inspection of Eq. \ref{spectralmodulation} shows that such a temperature variation of $\delta(\lambda,T)$ has an identical effect on the spectral modulation as a variation of $\phi_L$ (i.e. rotation of the linear polarization direction) has.
This thermal dependency of the multiple-order retarder therefore must be eliminated to unambiguously measure $\phi_L$.
Active thermal control is an option, although it increases the modulator's \mbox{(thermo-)}\-mechanical complexity and power consumption.
A second option is to create an ``athermal'' wave plate by combining two (or more) plates from different crystals with different thermo-optic constants, such that their thermal variations cancel (and their retardances don't).
Note that the measurement of $P_L$ is to first order unaffected by (residual) thermal variations of $\delta(\lambda,T)$.

\subsubsection{Design options}
To create an ``athermal'' multiple-order retarder, two different materials need to be found such that for a certain thickness ratio the residual thermal dependency of the combined retardance is minimal over the required wavelength range. 
The effective retardance of a single retarder plate is a strong function of incidence angle \cite{Keller2002}.
Preferably, for the crystal plate combination the FOV behavior is optimized.
Combinations of three or more crystal plates can be considered in order to gain additional degrees of freedom in the design space.
In the case of two plates $i=1,2$ with thicknesses $d_i$ and birefringence $n_{e,i}-n_{o,i}$, the combined retardance $\delta(\lambda,T)$ is given by:
\begin{equation}
\delta(\lambda,T) =  \delta_1(\lambda,T) \pm \delta_2(\lambda,T) = \Big|n_{e,1}(\lambda,T)-n_{o,1}(\lambda,T)\Big|\cdot d_1 \pm \Big|n_{e,2}(\lambda,T)-n_{o,2}(\lambda,T)\Big|\cdot d_2 \textrm{,}
\label{delta}
\end{equation}
with the $\pm$ sign for an adding/subtracting combination of crystals.
The absolute value of the birefringence is taken in order to describe the wave plates' alignment between their fast axes instead of their optic axes.
The requirement for combined athermal behavior is formulated as \cite{HaleDay,GuimondElmore}:
\begin{equation}
\gamma_1 \cdot \delta_1(\lambda_0) \pm \gamma_2 \cdot \delta_2(\lambda_0) = 0\textrm{,}
\label{athermaleq}
\end{equation}
with
\begin{equation}
\gamma_i = \frac{1}{\delta_i(\lambda_0)}\frac{\textrm{d}\delta_i(\lambda_0)}{\textrm{d}T} = \frac{1}{d_i}\frac{\textrm{d}d_i}{\textrm{d}T} + \frac{1}{n_{e,i}(\lambda_0)-n_{o,i}(\lambda_0)}\frac{\textrm{d}(n_{e,i}(\lambda_0)-n_{o,i}(\lambda_0))}{\textrm{d}T} \textrm{.}
\end{equation}
The vast majority of common crystals exhibit negative values of $\gamma_i$ \cite{HaleDay,thermooptic} and therefore a subtractive combination of crystals has to be chosen (minus sign of $\pm$ in Equations \ref{delta} and \ref{athermaleq}) to build an athermal retarder.

The FOV behavior of the crystal combination depends on the signs of $n_{e,1}-n_{o,1}$ and $n_{e,2}-n_{o,2}$.
Assuming that the first crystal has positive birefringence (e.g. quartz or MgF$_2$), the combined retardance for a subtractive combination as a function of a beam's inclination $\zeta$ and azimuth $\eta$ is given by \cite{Evans}:
\begin{eqnarray}
\delta(\lambda,\zeta,\eta) & =  & \delta_1(\lambda) \Bigg[1+ \frac{\zeta^2}{2n_{o,1}} \Bigg(\frac{\cos^2\eta}{n_{e,1}} -\frac{\sin^2\eta}{n_{o,1}}\Bigg)\Bigg] \ldots \nonumber\\
&\ldots & -\delta_2(\lambda) \Bigg[1+ \frac{\zeta^2}{2n_{o,2}} \Bigg(\frac{\sin^2\eta}{n_{e,2}} -\frac{\cos^2\eta}{n_{o,2}}\Bigg)\Bigg] \quad | \quad n_{e,2}-n_{o,2}>0\textrm{;}\nonumber\\
&\ldots & -\delta_2(\lambda) \Bigg[1+ \frac{\zeta^2}{2n_{o,2}} \Bigg(\frac{\cos^2\eta}{n_{e,2}} -\frac{\sin^2\eta}{n_{o,2}}\Bigg)\Bigg] \quad | \quad n_{e,2}-n_{o,2}<0\textrm{.}
\end{eqnarray}
The dependence of the retardance on the azimuth  $\eta$ can be made to (almost) disappear by an appropriate subtractive combination of crystals having both positive (or both negative) birefringence.
This effectively transforms the hyperbolical curves of equal retardance on the FOV into elliptical curves, which, on average, have a flatter behavior with the inclination.
However, the combination of crystals with opposite birefringence yields a larger FOV with relatively constant retardance for an anamorphic slit aligned $\pm45^\circ$ from the crystals' axes (i.e. in the $x$ or $y$ direction in Fig. \ref{schematic}.a) up to field angles larger than $\pm10^{\circ}$, whereas the acceptance cone half-angle is about 2$^\circ$.
Since the multiple-order retarder has to be aligned $\pm45^\circ$ from the axes of the quarter-wave retarder (see Fig. \ref{schematic}.a), a combination of a Fresnel rhomb or K-prism and the athermal multiple-order retarder consisting of opposite crystals with a slit spectrograph yields the most constant behavior along the slit.
Note that, in principle, a constant value of $\delta$ needs to be only guaranteed over the FOV that one spatial pixel within the spectrometer spans.
The response of the spectral modulator then needs to be calibrated for each spatial pixel independently.

The full Stokes spectral modulation method of Oka \& Kato \cite{OkaKato} would also significantly benefit from the use of athermal retarders.

\subsubsection{Thermal tests}
Using tabulated values for $\gamma_i$ from \cite{thermooptic}, we searched for an optimal combination of crystal that yields athermal behavior in the (extended) visible range from 350 to 800 nm.
We selected the common, robust crystals quartz, MgF${_2}$ and Al${_2}$O${_3}$ (sapphire) which all have high transmissions over a very large wavelength range (UV-NIR) and investigated the combinations of opposite crystals: quartz-sapphire and MgF${_2}$-sapphire.
The resulting thermal residuals over the wavelength range as a function of the thickness ratio $d_1:d_2$ were found to be minimal for a MgF${_2}$-sapphire combination with $d_1:d_2=2.7$.

However, the thermo-optic constants from the literature are too inaccurate for the exact determination of the thickness ratio for the athermal combination.
We therefore performed thermal tests on a combination of MgF${_2}$ and sapphire crystals with a variable thickness ratio.
A 1.1 mm thick sapphire crystal was mounted in a Soleil-Babinet mount from B. Halle in (subtractive) combination with a wedged MgF${_2}$ crystal with a thickness varying from 1.2 to 4.6 mm.
This set-up was positioned with the crystal axes at $\pm45^\circ$ between crossed Glan-laser polarizers and put in a thermal chamber.
Light from a halogen source outside the chamber was fed into the set-up in the chamber by means of an optical fiber.
A microscope objective and two diaphragms were used to produce a collimated beam of $\sim$1 mm diameter through the polarizers and the crystals, which is small enough to assume a constant thickness ratio over the beam.
Finally, another microscope objective feeds the light into a second fiber that is connected to an Ocean Optics USB2000 spectrograph (also outside the chamber) with a spectral range of 180-880 nm with 0.34 nm/pixel and a spectral resolution of 0.6 nm.
The temperature inside the chamber was varied from 5 to 50$^\circ$C in three equal steps for a number of different thickness ratios.

\begin{figure}[p]
\centerline{\includegraphics{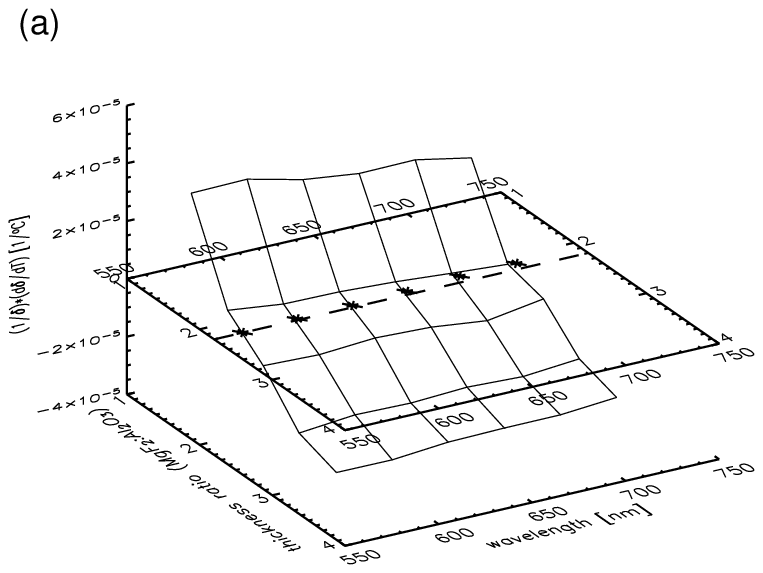}}
\centerline{\includegraphics{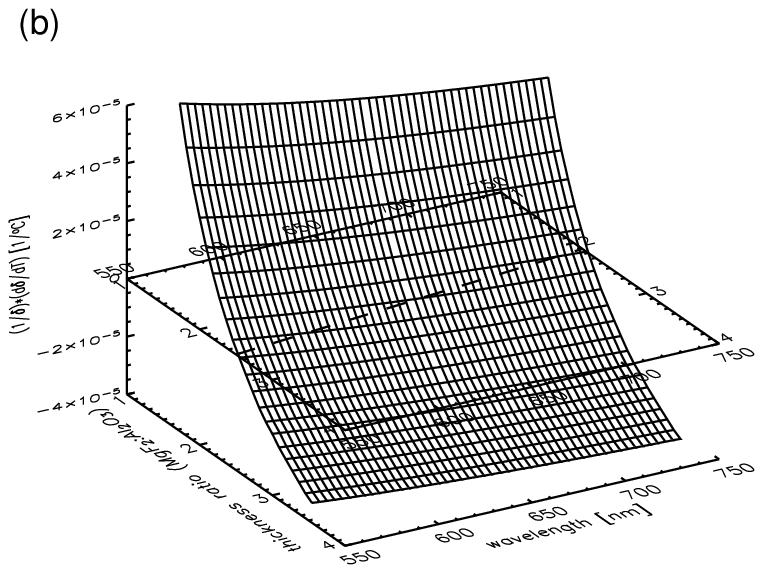}}
\caption{(a) Thermal test results for the combination of MgF$_2$ (with variable thickness) and sapphire as a function of wavelength (550-750 nm). Various thickness ratios (MgF$_2$:Al$_2$O$_3$) from 1.7 to 3.7 have been measured. The vertical axis represents the relative variation of the overall retardance of the crystal combination with temperature. The crosses represent the determined zero-crossings of the data for a certain wavelength. The dashed line represents the thickness ratio 1:2.2. (b) The corresponding theoretical curve with literature values for the thermo-optic constants. The dashed line here represents the zero-crossing of the curve. Snik\_AO105143\_fig3a.eps; Snik\_AO105143\_fig3b.eps}
\label{thermaltests}
\end{figure}

This set-up is essentially the same as depicted in Fig. \ref{schematic}.a.
The first polarizer that determines the input polarization to the modulator and its orientation is fixed and oriented in the same direction as the quarter-wave retarder would be.
This input polarization is an eigenvector of the quarter-wave retarder, which is therefore not implemented in this set-up.
Because $\phi_L$ is fixed, a measured variation of that angle is directly related to the temperature variation of the multiple-order retarder (see Eq. \ref{spectralmodulation}):
\begin{equation}
\phi_{L\textrm{\tiny{,meas.}}} = \phi_{L,0} + \frac{\pi}{\lambda}\frac{\textrm{d}\delta}{\textrm{d}T}\cdot \Delta T \textrm{.}
\label{phiT}
\end{equation}
The angle of linear polarization $\phi_L$ is retrieved by the algorithm presented in section \ref{retrieval}.
For a range of thickness ratios between MgF${_2}$ and sapphire and for wavelengths from 550 to 750 nm (where signal/noise was optimal), the variation of $\delta$ with temperature was obtained after linear fits of the data to Eq. \ref{phiT}.
The results, after averaging over 30 nm wavelength bins, are plotted in Fig. \ref{thermaltests}.a.
It is clear that the data cross the zero of the vertical axis which represents the thermal variation of the combined retardance.
The exact values for the thickness ratio for which the data cross zero are determined from second order polynomial fits to the data and are represented by the crosses.

For comparison, the theoretical curve based on thermo-optic constants from \cite{thermooptic} is plotted in Fig. \ref{thermaltests}.b.
The general shape of the data is reproduced, but the thickness ratio values for which athermal behavior is obtained differ by about 0.3.
The exact error bars on these results are difficult to estimate, but they are not very relevant to us as explained below.
Assuming that the inaccuracy of the literature values are the main cause for the discrepancy between the data and the theoretical curve, we adopt the results from the measurement data as the new best thickness ratio $d_1:d_2=2.4$.
It is clear from both the data as well as the theoretical curve that exact athermal behavior cannot be achieved for the entire wavelength range simultaneously.
The obtained thickness ratio therefore merely represents an optimum value for the wavelength range (350-800 nm).
At the edges of the wavelength range the residuals are less than 1.5$\times10^{-2}$ radians/$^\circ$C.
The situation can change slightly if the discrepancy between Fig. \ref{thermaltests}.a and b is due to (hard-to-judge) systematic errors in the determined thickness ratio.
In any case, the measurements prove that the set-up is perfectly athermal somewhere within the wavelength range.
The impact of thermal residuals on the modulator performance and a method to mitigate these effects are discussed in section \ref{errors}.

\subsection{Polarizer}
Many options exist for the choice of the polarizer or polarizing beam-splitter, mostly depending on the wavelength range.
The extinction ratio should be as large as possible, hence crystal polarizers such as Glan or Wollaston prisms are preferred.
If the spectral modulator is located in the entrance pupil of the instrument and is based on a dual beam system, a Wollaston followed by one or two objective lenses forms two separated images of the sources.
If the spectrometer is a slit spectrograph, the separation by the Wollaston prism needs to be parallel to the slit direction.

%=======

\section{Retrieval algorithm}\label{retrieval}
Since the spectral modulator entangles the linear polarization information in the intensity spectrum, a dedicated algorithm was developed to retrieve all relevant information.
We assume that $P_L$ and $\phi_L$ can vary randomly with wavelength.
Together with the facts that the modulation is only periodic with the inverse wavelength and that $\delta$ varies with wavelength, this seriously complicates retrieval methods based on Fourier or wavelet analyses or on a Hilbert transform.
Instead, we fit Eq. \ref{spectralmodulation} to a range of narrow wavelength windows.
The size of these windows should be larger than one modulation period and roughly determines the final spectral resolution.
The algorithm is slightly different for a single beam and a dual beam system (i.e. without or with a polarizing beam-splitter), because the intensity spectrum is trivially obtained in the second case by adding the two beams.
We choose not to assume any knowledge about the physical parameters of the spectral modulator and the  spectrometer, because the error propagation from those parameters is sometimes difficult to judge.
All non-ideal effects are to be taken into account by calibration (see section \ref{errors}).
A first estimate for the dispersion of the retardance $\delta(\lambda)$ is supplied to the retrieval algorithm, but this estimate can be improved after calibration.

The basic steps of the algorithm are:
\begin{enumerate}
\item{} Determination of the spectral windows. For each wavelength value measured by the spectrometer, a spectral window equal to the local modulation periodicity $\Delta \lambda$ is determined as follows for a certain order $k$ of the multiple-order retarder:
\begin{equation}
\delta=k\lambda \approx (k-\frac{1}{2})(\lambda+\frac{\Delta\lambda}{2}) \approx (k+\frac{1}{2})(\lambda-\frac{\Delta\lambda}{2}) \textrm{;}
\end{equation}
\begin{equation}
\Delta\lambda=\frac{\lambda^2}{\delta(1+\lambda^2/4\delta^2)} \textrm{.}
\label{windows}
\end{equation}
The modulation periodicity therefore has a quadratic behavior for $\delta \gg \lambda$.
At the edges of the wavelength range, no windows can be created that completely contain measurement values, so some data is lost there, especially on the red side of the spectral range where the modulation period is much longer.
\item{} For the single beam system the intensity spectrum at the center of each spectral window is estimated as the average value of the data within the spectral window. For the dual beam system the intensity spectrum is obtained by adding both modulated spectra.
\item{} The measured spectrum is divided by the obtained intensity spectrum.
\item{} A first estimate for $P_L$ is obtained by determining the minimum and maximum value of the normalized data within each window.
\item{} An artificial reference signal is created with $\phi_L=0$ and $P_L=1$.
\item{} A first estimate for $\phi_L$ in each window is obtained after cross-correlation of the normalized signal and the reference signal and a $\cos^2(\lambda+\lambda_{\phi})$ fit of the cross-correlation signal. The determined phase shift $\lambda_{\phi}$ between the data and the reference signal is transformed into an estimate of the angle of linear polarization:
\begin{equation}
\phi_{L,\textrm{\tiny{est.}}}=\frac{\pi\cdot \lambda_{\phi}}{\Delta\lambda(\lambda)}
\end{equation}
\item{} With the initial estimates for $P_L$ and $\phi_L$, full curve-fitting of the normalized data within each window is performed (cf. Eq. \ref{spectralmodulation}). Because the solution can run away with $\phi_l$, we apply periodic constraints such that $0\leq \phi_L \leq \pi$ and $0\leq P_L \leq 1$. The values of $\delta(\lambda)$ are calculated from the known crystal thicknesses and literature values for the birefringence \cite{thermooptic}.
\item{} The final results for $P_L(\lambda)$ and $\phi_L(\lambda)$ are smoothed with the window widths of Eq. \ref{windows} because information at higher spectral resolution cannot be determined reliably.
\end{enumerate}

\begin{figure}[p]
\centerline{\includegraphics{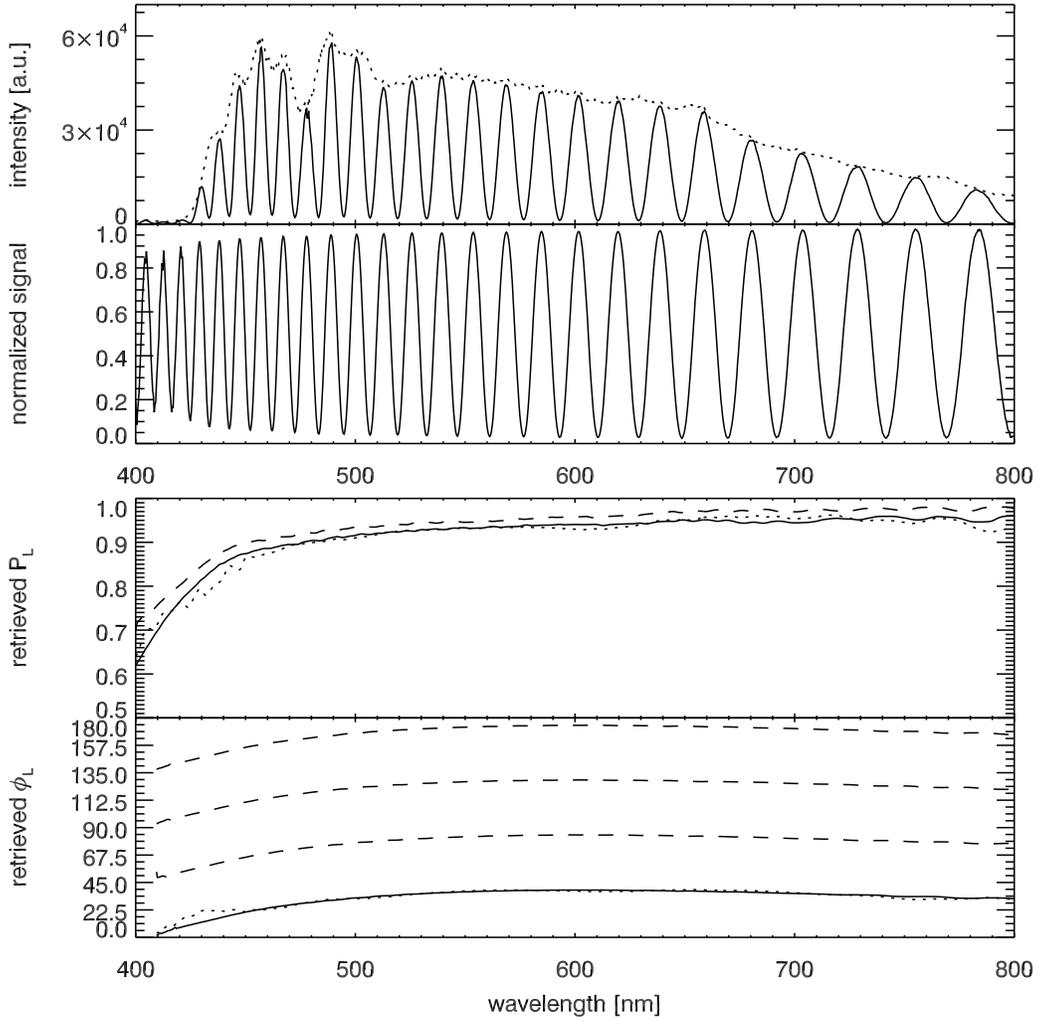}}
\caption{The upper panel shows a modulated spectrum obtained with the prototype described in the text (solid curve). The dotted curve represents the unmodulated intensity spectrum obtained after adding the signals from two orthogonal orientations of the polarizer. The normalized modulation signal is shown below. The lower panel shows the results for $P_L(\lambda)$ and $\phi_L(\lambda)$ as obtained by the retrieval algorithm. Offsets from the ideal values of  $P_L(\lambda)$ (=1) and $\phi_L(\lambda)$ (constant) are clearly observed. The solid curve corresponds to the signal shown in the upper panel using the dual beam method, whereas the dotted curve shows the results from the single beam method ($\phi_L$ is the same in both cases). The dashed curves represent rotations of the input polarization in steps of 45$^\circ$, which are reproduced by the $\phi_L$ measurements. The dashed curve for $P_L(\lambda)$ represents the measurement with the input polarization direction coinciding with the K-prism axis, for which the measurement efficiency is significantly higher than for the situation with input at 45$^\circ$ from that (solid curves). Snik\_AO105143\_fig4.eps}
\label{calibrationdata}
\end{figure}

Experimental results obtained with a prototype spectral modulator consisting of a fused silica K-prism, a subtractive combination of 2.53 mm of MgF$_2$ and 1.10 mm of sapphire, and a rotatable Glan-laser polarizer in combination with the fiber-fed 0.6 nm resolution spectrograph mentioned above are presented in Fig. \ref{calibrationdata} as well as the retrieved values of $P_L(\lambda)$ and $\phi_L(\lambda)$.
The polarizer was used to mimic a dual beam system by rotating it back and forth by 90$^\circ$.
The retrieved values using the single beam method (dotted curves in the lower panel) are very similar to those of the dual beam method (solid curves).
The input polarization was created with another rotatable Glan-laser polarizer, and therefore $P_L$ should be equal to 1 and $\phi_L$ should be constant.
A rotation of the input polarizer leads to the correct increase of the retrieved $\phi_L(\lambda)$.
However, it is obvious that some deviations are observed in the measured values of both $P_L(\lambda)$ and $\phi_L(\lambda)$, and therefore these measurements represent a calibration of non-ideal effects (see next section).

Using simulated spectral data in combination with the retrieval algorithm, we determined the optimum value of $\delta$ for a required final spectral resolution of 20 nm using a spectrometer with 2 nm resolution from 350-800 nm. 
The minimum value for the squared difference between simulated and retrieved data was found for $\delta \approx 25 \mu$m. 
The accuracy with which the simulated data was retrieved was within 5\% relative for $P_L$ and within $\pm10^\circ$ for $\phi_L$ (see \cite{aerosols}). 
It is difficult to judge what the ultimate sensitivity (i.e. the noise level of the $P_L$ measurement) and accuracy of the spectral modulator can be, since this is likely dominated by unforeseen instrumental effects. 
Further measurements with (and calibrations of) the spectral modulator prototype are required to investigate this.

%=======

\section{Error sources and calibration}\label{errors}
A number of instrumental error sources can be identified \textit{a priori}.
We present a list of potential major error sources and ways to mitigate those.
Fortunately the measurement principle is quite independent for $P_L$ and $\phi_L$ so that most errors influence only one of these measurements.

\subsection{Static errors}
Many alignment errors and static, non-ideal properties of the components of the spectral modulator can lead to a reduced efficiency of the $P_L$ measurements.
These efficiencies can depend on $\phi_L$ as shown in the lower panel of Fig. \ref{calibrationdata}.
They are also likely to vary across the FOV.
All these effects are assumed to be stable in time and are calibrated for each pixel in the FOV by a measurement like in Fig. \ref{calibrationdata} as elaborated in section \ref{calibration}.
A number of effects can influence the $P_L$ measurement:
\begin{itemize}
\item{} Opto-mechanical alignment errors can influence the measurement efficiency in many ways. Individual alignment sensitivities can be assessed through the Mueller matrix calculus.
\item{} An imperfect quarter-wave retarder results in an incomplete translation of the $(Q/I, U/I)$ plane to the $(Q/I, V/I)$ plane (see Fig. \ref{schematic}.b). The modulation amplitude decreases, and therefore linear polarization in the $\pm U$ direction is measured with less efficiency. This effect is clearly visible in the data shown in Fig. \ref{calibrationdata}.
\item{} Pleochroism (partial polarizibility) in (one of) the multiple-order retarder crystals reduces the modulation amplitude, but this depends on $\phi_L$. An extinction ratio $E$ of the combined multiple-order retarder (which should be close to 1) leads to a maximum reduction of the measured $P_L$ of $2\sqrt{E}/(1+E)$.
\item{} An imperfect polarizer also directly reduces the modulation amplitude.
\item{} The modulation amplitude is also reduced by convolution of the spectra to be measured with the spectrograph profile. This effect is more severe in the blue side of the spectrum as is seen in Fig. \ref{calibrationdata}.
\item{} Variation of $\delta$ over a spatial pixel or over the beam footprint on the multiple-order retarder leads to a smearing of modulations with slightly different periodicities, which leads to an effective reduction of the modulation amplitude.
\end{itemize}

The zero point of the $\phi_L(\lambda)$ measurement can exhibit offsets due to the following causes:
\begin{itemize}
\item{} Limited opto-mechanical alignment accuracy.
\item{} The actual value of $\delta(\lambda)$ is not perfectly well known. This influences the retrieval of $\phi(\lambda)$ as shown in the lower panel of Fig. \ref{calibrationdata}. Assuming that $\delta(\lambda)$ is constant in time and that the same retrieval algorithm is used for both the calibration and for the actual data, the calibration determines the zero point of $\phi_L$ as a function of wavelength. It is expected that this effect also influences the retrieval efficiency of $P_L$, but this is then consistently calibrated in the same manner. The retrieved trend of $\phi_L(\lambda)$ could be used to establish the exact values of $\delta(\lambda)$, such that $\phi_L(\lambda)$ becomes flat.
\end{itemize}

\subsection{Calibration}\label{calibration}
The static errors listed above can all be calibrated by controlling the input polarization (preferably the degree and the angle of linear polarization) and analyzing the output of the spectral modulator.
The most straightforward calibration routine consists of positioning a rotating (perfect) polarizer in front of the spectral modulator.
The retrieved values of $P_L(\lambda)$ and $\phi_L(\lambda)$ represent the reference data for any other observation  with the spectropolarimeter.
Since the degree of linear polarization measurement potentially depends on the angle, the efficiency of $P_L(\lambda)$ needs to be mapped as a function of $\phi_L(\lambda)$.
Care has to be taken that the calibration beam does not have a Stokes $V$ component, since an imperfect spectral modulator may be sensitive to circular polarization.

The calibration set-up with a rotating polarizer is not able to vary the degree of linear polarization, which is desirable since many spectropolarimeters are designed to measure signals with $P_L \ll 1$.
Several (non-linear) instrumental effects that hamper the accurate measurement of low degrees of polarization may not be properly calibrated with an input $P_L = 1$.
Also, the sensitivity of the polarimeter can thus be established and perhaps improved by identifying limiting effects.
A calibration set-up which allows for full control of $1 \leq P_L \leq 0$ and $\phi_L$ can be constructed using a `theta cell' \cite{thetacell1,thetacell2,snikthetacell}.
In any case, the accuracy of the calibration of a spectral modulator should be at least 5\% of $P_L$ and $\pm10^\circ$ in $\phi_L$, because the noise levels in the retrieved values of $P_L$ and $\phi_L$ are easily suppressed below those values.

\subsection{Dynamic errors}
A number of dynamic effects can modify the performance of a spectral modulator in time.
The influence of these effects could be detectable in the data:
\begin{itemize}
\item{} Bias drift and variable stray light with a measured total amount $s_d$ lead to a reduction of $P_L$ by $s_0/(s_0+2s_d)$. It is therefore crucial to use a detector with negligible dark drift and to deal with stray light by adequate baffling.
\item{} The transmission ratio $t_1(\lambda)/t_2(\lambda)$ between the two beams can slightly vary over time e.g. due to aging of optical coatings and/or the collection of dust. This effect could be detected in the intensity spectrum obtained by adding the two signals. A Fourier analysis of this spectrum allows for the detection of residual spectral modulation and redetermination of  $t_1(\lambda)/t_2(\lambda)$ such that the Fourier power of the modulation disappears. An alternative method is based on the fact that the RMS of the obtained intensity spectrum should be minimal for the correct value of  $t_1(\lambda)/t_2(\lambda)$.
\item{} The ``athermal'' multiple-order retarder always has some residual temperature sensitivity at least in parts of the wavelength range. These thermal effects, to first order, only influence the $\phi_L$ measurement, in a predictable way. However, if the $P_L$ measurements are affected by non-ideal effects depending on $\phi_L$, the angle needs to be determined accurately to apply the calibration to the measurements. A correction can be made if the temperature of the modulator during the measurement is known. An indirect correction method can be devised in the case that (part of) the measurable $\phi_L(\lambda)$ is constant (e.g. in cases of single scattering). The deviations due to the thermal effects can then be easily identified and corrected for over the entire wavelength range.
\end{itemize}

%=======

\section{Acknowledgments}
The authors thank two anonymous reviewers for their constructive comments on the manuscript. This research was conducted within the framework of a PEP grant from the Dutch Agency for Aerospace Programs (NIVR). We thank the technical staff of Dutch Space for their assistance with the thermal tests. The calibration measurements were performed by Gerard van Harten. A provisional patent application for the passive, athermal spectral modulator for full linear spectropolarimetry has been filed by the Utrecht University Holding.

\newpage
\textbf{Figure caption listing page}

Fig. 1. (a) Schematic set-up of the spectral modulator. The solid and dashed lines of the retarders represent the (orthogonal) fast and slow axis respectively. (b) Illustration of the spectral modulation principle on the Poincar\'e sphere.
\newline\\
Fig. 2. Design of a retroreflecting quarter-wave retarder prism based on 3 TIRs in a fused silica rhomb. The path of light with an angle of 2$^\circ$ from normal incidence is indicated in light grey. 
\newline\\
Fig. 3. (a) Thermal test results for the combination of MgF$_2$ (with variable thickness) and sapphire as a function of wavelength (550-750 nm). Various thickness ratios (MgF$_2$:Al$_2$O$_3$) from 1.7 to 3.7 have been measured. The vertical axis represents the relative variation of the overall retardance of the crystal combination with temperature. The crosses represent the determined zero-crossings of the data for a certain wavelength. The dashed line represents the thickness ratio 1:2.2. (b) The corresponding theoretical curve with literature values for the thermo-optic constants. The dashed line here represents the zero-crossing of the curve.
\newline\\
Fig. 4. The upper panel shows a modulated spectrum obtained with the prototype described in the text (solid curve). The dotted curve represents the unmodulated intensity spectrum obtained after adding the signals from two orthogonal orientations of the polarizer. The normalized modulation signal is shown below. The lower panel shows the results for $P_L(\lambda)$ and $\phi_L(\lambda)$ as obtained by the retrieval algorithm. Offsets from the ideal values of  $P_L(\lambda)$ (=1) and $\phi_L(\lambda)$ (constant) are clearly observed. The solid curve corresponds to the signal shown in the upper panel using the dual beam method, whereas the dotted curve shows the results from the single beam method ($\phi_L$ is the same in both cases). The dashed curves represent rotations of the input polarization in steps of 45$^\circ$, which are reproduced by the $\phi_L$ measurements. The dashed curve for $P_L(\lambda)$ represents the measurement with the input polarization direction coinciding with the K-prism axis, for which the measurement efficiency is significantly higher than for the situation with input at 45$^\circ$ from that (solid curves).

\end{document}